\documentclass[twocolumn,floatfix,amsfonts,amssymb,notitlepage]{revtex4-1}
\usepackage{color}
\usepackage{graphicx}
\usepackage{amsmath}
\usepackage{latexsym}
\usepackage{epstopdf}
\usepackage{amsmath}
\usepackage{amssymb}
\usepackage{dsfont}
\usepackage{multirow}
\usepackage{mathtools}
\usepackage{tcolorbox}

\newcommand{\beq}{\begin{equation}}
\newcommand{\eeq}{\end{equation}}
\newcommand{\bea}{\begin{eqnarray}}
\newcommand{\eea}{\end{eqnarray}}

\newcommand{\bebox}{\begin{tcolorbox}}
\newcommand{\eebox}{\end{tcolorbox}}

\newcommand{\eq}{\begin{equation}}
\newcommand{\en}{\end{equation}}
\newcommand{\ear}{\begin{eqnarray}}
\newcommand{\rae}{\end{eqnarray}}
\newcommand{\Z}{\mathbb{Z}}
\newcommand{\C}{\mathbb{C}}

\newcommand{\id}{\mathds{1}}
\newcommand{\be}{\begin{eqnarray}}
\newcommand{\ee}{\end{eqnarray}}
\newcommand{\non}{\nonumber}

\makeindex

\begin{document}
\title{Integrable quantum spin chains with free fermionic and parafermionic spectrum}

\author{Francisco C. Alcaraz}
\email{alcaraz@ifsc.usp.br}
\affiliation{ Instituto de F\'{\i}sica de S\~{a}o Carlos, Universidade de S\~{a}o Paulo,
Caixa Postal 369, 13560-970, S\~{a}o Carlos, SP, Brazil}

\author{Rodrigo A. Pimenta}
\email{pimenta@ifsc.usp.br}
\affiliation{ Instituto de F\'{\i}sica de S\~{a}o Carlos, Universidade de S\~{a}o Paulo,
Caixa Postal 369, 13560-970, S\~{a}o Carlos, SP, Brazil}
\date{\today{}}

\begin{abstract}
We present a general study of the large family of exact integrable quantum 
chains with multispin interactions introduced recently in \cite{AP2020}.
The exact integrability follows from the algebraic properties of the 
energy density operators defining the quantum chains. The Hamiltonians are 
characterized by a parameter $p=1,2,\dots$ related to the  number of 
interacting spins in the multispin interaction. 
In the general case the quantum spins are of infinite dimension. In special 
cases, characterized by the parameter $N=2,3,\ldots$, the quantum chains 
describe the dynamics of  $Z(N)$ quantum spin chains. 
 The simplest case $p=1$ 
corresponds to the free fermionic quantum Ising chain ($N=2$) or the 
$Z(N)$ free parafermionic quantum chain. The eigenenergies of the quantum 
chains are given in terms of the roots of special polynomials, and for general 
values of $p$ the quantum chains are characterized by a free fermionic ($N=2$) or 
free parafermionic ($N>2$) eigenspectrum. The models have a special critical 
point when all coupling constants are equal. At this point the 
ground-state energy is exactly calculated in the bulk limit, and our analytical and 
numerical analyses indicate that the models belong to universality classes 
of critical behavior with dynamical critical exponent $z = (p+1)/N$ and 
specific-heat exponent $\alpha = \max\{0,1-(p+1)/N\}$.
\end{abstract}

\maketitle

\section{Introduction} 

The study of free fermionic systems like the Ising model
in a transverse field or
the quantum XY model has proven to be an important
step towards
the understanding of quantum many body interacting systems
\cite{Schultz:1964fv,Pfeuty}.
Since the entire spectrum of these models is obtained exactly
in a finite geometry they provide the ideal framework
for the study of new
mathematics and physics in Condensed Matter, Statistical Physics
and Quantum Information Theory.

Extensions of the models
with a free eigenspectrum have been also introduced. One example
is the $Z(N)$ generalization of the Ising model
in a transverse field introduced by 
Baxter \cite{Baxter:1989bma,Baxter:1989vv} in which
the eigenspectrum is described by free parafermions
\cite{Fendley:2013,Baxter2014,Au_Yang_2014,auyang2016parafermions,
Alcaraz_2017,Alcaraz_2018,Liu_2019}.
More recently, a new free fermionic $Z(2)$ model characterized by a three-spin
interaction was discovered by Fendley \cite{Fendley:2019}. The work
\cite{Fendley:2019} motivated us for the discover  of a large family of
exactly integrable quantum spin chains. 
In general, the associated quantum spin chains act on a infinite dimensional 
vector space, even 
in a finite lattice. However at special cases the dimension is truncated and the models
turn out describing $Z(N)$ quantum spin chains with multispins interactions \cite{AP2020}.

The eigenspectrum of this new  family of models has a
free fermionic ($N=2$) or free parafermionic ($N>2$) nature. 
The solution of their free-particle eigenspectra cannot be obtained by the standard 
Jordan-Wigner
transformation. As we show in this paper, the simplest way 
to solve their eigenspectra  is by noticing that
the general Hamiltonians are based on representations of a simple
exchange algebra. The algebra defining the Hamiltonians has an integer
parameter $p\geq 1$ and its representations turn out to be
quantum chains with $(p+1)$-multispin interactions.

The exchange algebra allows us the construction
of a set of mutually commuting charges, including the Hamiltonian.
The solution of the spectral problem, in the $Z(N)$-truncated cases, 
is obtained thanks to a product formula (or inversion
relation for $N=2$)
satisfied by the generating function of the charges. This
relation enables us to express the quasienergies
of the free particles in terms of the roots of a family
of polynomials. The polynomials, and consequently
the eigenspectrum of the quantum chains, follow solely from the algebraic properties
of the energy density operators defining the quantum chain.
This imply that a given set of pseudoenergies describes the eigenspectrum
of distinct quantum chains associated with different representations
of the exchange algebra.

The commutativity of the charges built from the exchange algebra
has been proved for $p=1$ and arbitrary $N$ in \cite{Fendley:2013} and
for $N=p=2$ in \cite{Fendley:2019}. 
In the communication \cite{AP2020} we announced the commutation for the 
truncated model with arbitrary values of $p$ and $N$. In this paper we 
prove the announced commutation in the more general case where the Hilbert 
space associated to the quantum chain is not necessarily truncated.

In the $Z(N)$-truncated models the
product formula, or inversion relation, satisfied by the generating function
is similar to the one satisfied by the transfer matrix of the $\tau_2$ model
\cite{Bazhanov:1989nc,Baxter:1999mn,Tarasov:1991mf,Baxfunc}, which is associated 
with the case where $p=1$. For $p=N=2$,
the inversion
relation has been proved in \cite{Fendley:2019}. 
In \cite{AP2020} we conjectured the inversion relation for general $p$ and 
general $Z(N)$-truncated cases. In this paper we prove this conjecture for
 $N=2$ and $N=3$, and general values of $p$.
Our method is solely based on a recurrence relation
satisfied by the generating function and can be
extended for $N>3$, case by case.

The polynomials determined by the product formula
play a crucial role in the physical analysis of the Hamiltonians.
The fact that the pseudoenergies
of the free particles are related to the roots of the polynomials
allows the evaluation of the eigenspectrum for quite large  system sizes. 

The quantum chains for general values of $N$ and $p$ have a special
critical point. The mass gap and specific heat calculations, at 
the critical point, allow us the calculation of the dynamical 
critical exponent $z$ analytically and the specific-heat exponent 
$\alpha$ numerically.

We finally remark that the diagonalization
of the Hamiltonians are performed independently of
the representation of the exchange algebra. Nevertheless,
once a representation is chosen, extra degeneracies can appear in the
eigenspectrum.

This paper is organized as follows. In Section \ref{sec:h}
we introduce the general family of integrable quantum
Hamiltonians with the algebraic properties defining the energy
density operators. We also present several representations
of these models making connections with already known models.
In Section \ref{sec:integrability} we prove that the Hamiltonians
belong to a family of mutually commuting operators. Next, in Section
\ref{sec:prod} we prove the inversion relation ($N=2$) and the product
formula ($N=3$) for arbitrary $p$. The polynomial fixing the
quasienergies is considered in Section \ref{sec:pol}.
In Section \ref{sec:crit} we study some of the physical properties
of the multispin Hamiltonians. In Section \ref{sec:adic} some additional
charges for specific representations are constructed. Our conclusions
and further directions of investigation are given in Section \ref{sec:conc}.
Finally in Appendices A and B, we present some technical details for the 
proof of Sec. \ref{sec:pol}, and an example of application to a small 
quantum chain, respectively.

\section{The integrable quantum chains}\label{sec:h}

The integrable models we construct are defined in terms
of $M$ generators $h_i$ ($i=1,\ldots,M$):
\be\label{Hgen}
\mathcal{H}= -\sum_{i=1}^Mh_i \,.
\ee
The generators satisfy a simple exchange algebra
characterized by an integer parameter $p=1,2,\dots$  given by,
\be \label{halgebra1}
h_i h_{i+m} &=& \omega h_{i+m}h_i
\quad \textrm{for}\quad 1\leq	m \leq p\,, \nonumber \\
\left[h_i, h_j\right] &=& 0 \quad \textrm{for} \quad |i-j|>p .
\ee
where $\omega$ is a general complex c-number. As we show in Section \ref{sec:integrability},
the relations (\ref{halgebra1})
imply the exact integrability of (\ref{Hgen})
since we can construct a set of mutually commuting
charges. This set is infinite in the bulk limit $M\rightarrow \infty$.
  
We define a word in the algebra by an arbitrary product of letters 
(generators) $h_1^{s_1}h_2^{s_2}\cdots h_M^{s_M}$, where 
$s_i=0,1,2,\ldots$ .
 It is important to stress that relations
(\ref{halgebra1}) 
does not fix any power $h_i^a$ of the generators
and consequently
the number of independent words of the algebra
is infinite.
In order
to have a finite number of words $(N^M)$ we can include besides
(\ref{halgebra1})  a closure relation,
as for example
\be\label{halgebra3}
h_i^N = \lambda_i^N\,,\quad \omega = e^{2i\pi/N}\, ,
\ee
where $\lambda_i^N$ is a c-number and 
 $N=2,3,\dots$. When (\ref{halgebra3}) is taken into account together
with relations (\ref{halgebra1}), we are going to show that 
(\ref{Hgen}) has a free fermionic ($N=2$)
or free parafermionic ($N>2$) eigenspectrum, \textit{i.e.},
the eigenenergies $E^{\{s_i\}}$ satisfy 
\be\label{EHgen}
-E^{\{s_i\}} = \omega^{s_1} \epsilon_1 +
\omega^{s_2} \epsilon_2 + \cdots + \omega^{s_{\overline{M}}} \epsilon_{\overline M}\,,
\ee
where we define
\be\label{bar}
\overline{M} \equiv \mbox{int}\left(\frac{M+p}{p+1}\right)=\Big{\lfloor}{\frac{M+p}{p+1}}\Big{ \rfloor}
\ee
and
$s_i\in\{0,1,\dots,N-1\}$. The pseudoenergies $\epsilon_i$ ($i=1,\dots,\overline M$)
are given by the roots of special polynomials, as we show in Sec. \ref{sec:prod}.

Let us remark that the algebra (\ref{halgebra1})-(\ref{halgebra3}) for $p=1$
has been shown to be important in the study of generalized Clifford algebra
\cite{Morris1967,Morris1968,Mittag,Martin1989,Truong1986,K1986,Ramakrishnan1986,Jaffe:2014ama}.

While the diagonalization of (\ref{Hgen}) in the truncated models, is performed using only
(\ref{halgebra1})-(\ref{halgebra3}), representations of the algebra
lead to interesting
quantum spin chains.

A simple representation of (\ref{halgebra1})-(\ref{halgebra3}) with $N=2$ and 
$p=1$
for  odd values of $M$ is given in terms of the spin-$1/2$ Pauli matrices
$\sigma_i^{x,z}$:
\be
&&h_{2i-1} = \lambda_{2i-1}\sigma_i^x\quad
\textrm{for}\quad i =1,\dots,\frac{M+1}{2}\,,\non\\
&&h_{2i} =  \lambda_{2i}\sigma_i^z\sigma_{i+1}^z\,
\quad \textrm{for}\quad i =1,\dots,\frac{M-1}{2} \,,
\ee
leading to the Hamiltonian,
\be\label{HIsing}
\mathcal{H}_I =
-\sum_{i=1}^{L}\lambda_{2i-1}\sigma_i^x
-\sum_{i=1}^{L-1}
\lambda_{2i}\sigma_i^z\sigma_{i+1}^z\,,
\ee
also known as the free fermionic quantum Ising chain with $L=\frac{M+1}{2}$ sites
and coupling constants $\{\lambda_i\}$.

Similarly, for $p=1$, arbitrary $N$ and odd $M$, a representation of 
 (\ref{halgebra1})-(\ref{halgebra3}) is given in terms of the
$Z(N)$ generalizations of the $N \times N$ Pauli matrices satisfying,
\be\label{Weyl}
XZ = \omega ZX\,, ~ X^N = Z^N = 1\,, ~ Z^\dagger = Z^{N-1}\,,
\ee
and
\be
h_{2i-1} =
\lambda_{2i-1}X_i\quad &\textrm{for}&\quad i =1,\dots,\frac{M+1}{2}\,,\non\\
h_{2i} = 
\lambda_{2i}Z_i^\dagger Z_{i+1}\, \quad &\textrm{for}&\quad i =1,\dots,\frac{M-1}{2} \,.
\ee
The Hamiltonian (\ref{Hgen}) in this case,
\be\label{HBaxter}
\mathcal{H}_B = -\sum_{i=1}^{L}
\lambda_{2i-1}X_i-\sum_{i=1}^{L-1}
\lambda_{2i}Z_i^\dagger Z_{i+1}\,,
\ee
reproduces the free parafermionic Baxter $Z(N)$ model
with $L=\frac{M+1}{2}$ sites, introduced in \cite{Baxter:1989bma,Baxter:1989vv}.

In the case where $M$ is even, $N=2$ and $p=1$,
an interesting representation of (\ref{Hgen})
is given by,
\be
&&h_{2i-1} = \lambda_{2i-1}\sigma_i^x\quad
\textrm{for}\quad i =1,\dots,\frac{M-2}{2}\,,\non\\
&&h_{2i} =  \lambda_{2i}\sigma_i^z\sigma_{i+1}^z\,
\quad \textrm{for}\quad i =1,\dots,\frac{M-2}{2} \,,\non\\
&&h_{M-1} = \lambda_{M-1}\sigma_{M/2}^x\,,
\quad h_{M} = \lambda_{M}\sigma_{M/2}^z\,,
\ee
and
\be\label{HIsingeven}
\mathcal{H}_{\mbox{\scriptsize{imp}}} =
 -\sum_{i=1}^{L-1}\left(\lambda_{2i-1}\sigma_i^x
+\lambda_{2i}\sigma_i^z\sigma_{i+1}^z\right)-\vec{S}_0\cdot
\vec{\sigma}_{L}\,,
\ee
that represents a quantum Ising chain, with $L=M/2$ sites, interacting
at one of its ends with a magnetic impurity with
components $\vec{S}_0 = (S_0^x,S_0^z)=(\lambda_{M-1},\lambda_M)$.

The simplest representation of (\ref{halgebra1})-(\ref{halgebra3}) with
$N=2$ and $p=2$ recovers the free-fermionic three-spin interaction
Hamiltonian,
\be\label{HFendley}
\mathcal{H}_F = -\sum_{i=1}^{L-2}
\lambda_{i}\sigma_i^z\sigma_{i+1}^x\sigma_{i+2}^x\,.
\ee
introduced in \cite{Fendley:2019}.
A general example of a free fermionic ($N=2$) or
free parafermionic ($N>2$) with arbitrary values of $p=1,2,\ldots$
is given by \cite{AP2020},
\be\label{HP}
\mathcal{H}_P = -\sum_{i=1}^M h_i = -\sum_{i=1}^M
\lambda_i Z_i Z_{i+1}\cdots Z_{i+p-1} X_{i+p}.
\ee

A general interesting representation of (\ref{halgebra1})-(\ref{halgebra3}),
 that we call 
{\it{word representation}}
is the one where the generators 
act on a vector space spanned by the basis $\{|s_1,\ldots,s_M>\}$ with a 
biunivocal correspondence with the $N^M$ independent words formed by the 
normalized product of generators 
\be \label{words}
|s_1,\ldots,s_M> \leftrightarrow (h_1/\lambda_1)^{s_1}\cdots
(h_M/\lambda_M)^{s_M},
\ee
where $s_i=0,1,\ldots,N-1$. The action of $h_i$ ($i>p$) in this basis gives 
\bea \label {word1} \nonumber
&&h_i|s_1,\dots,s_M> \leftrightarrow \Omega 
\lambda_i |s_1,\ldots,s_{i-1},s_i^+, s_{i+1},\ldots,s_M>, \nonumber \\
&&\mbox{with } \Omega= \omega^{\sum_{j=1}^p s_{i-j}} \mbox{ and }
s_i^+= s_i+1, \textrm{mod}\, N. \nonumber 
\ee
 From the properties of the $Z(N)$ 
matrices ($Z_j,X_j$) we can identify
\be \label{wordrepp}
 h_i = \begin{cases} 
    \lambda_i\left(\prod_{j=1}^{i-1} Z_j \right) X_i, & \mbox{if } 1\leq i \leq p ;\\
    \lambda_i\left(\prod_{j=i-p}^{i-1} Z_j\right) X_i, & \mbox{if } p+1\leq i\leq  M, \end{cases} 
\ee
in the basis $|\tilde{s}_1,\ldots,\tilde{s}_M>$ where the $\{Z_i\}$ are 
diagonal, i. e., 
\be \label{znalgebra}
Z|\tilde s_i\rangle = e^{2i\pi\tilde s_i/N }|\tilde s_i\rangle\,,\quad
X|\tilde s_i\rangle = |\tilde s_i+1, \textrm{mod}\,N\rangle\,.
\ee

The Hamiltonian is then given, in this representation, by the $Z(N)$
quantum chain with multispin interactions:
\be\label{HA}
&&\mathcal{H}_A = -\sum_{i=1}^p \lambda_i \left(\prod_{j=1}^{i-1}
Z_j\right) X_i-\sum_{i=p+1}^M \lambda_i
\left(\prod_{j=i-p}^{i-1} Z_{j}\right)X_i\,.\non\\
\ee
In the particular case $N=2$ and $p=1$ we have the simple
nearest-neighbor interacting Hamiltonian,
\be\label{HA1}
\mathcal{H}_A^{(p=1)} = -\lambda_1 \sigma_1^x-
\sum_{i=2}^M \lambda_i \sigma_{i-1}^z\sigma_i^x\,.
\ee
It is important to observe that the Hamiltonians (\ref{HBaxter},\ref{HP},\ref{HA})
are Hermitian in the fermionic cases ($N=2$) and non-Hermitian in
the parafermionic cases ($N>2$).

\section{Exact Integrability and Conserved Charges}\label{sec:integrability}

In this Section we show that the
general Hamiltonian (\ref{Hgen}) given in terms
of the generators $\{h_i\}$ of the exchange algebra (\ref{halgebra1}),
is part of a set
of commuting operators (charges), which became infinite in the bulk limit ($M\rightarrow\infty$), being exactly integrable.

The conserved charges follow directly from the algebraic rules 
(\ref{halgebra1}).
It is not necessary the closure relation (\ref{halgebra3}).
A given charge $\ell$ ($\ell=0,1,\dots,M$) is obtained
by summing all the products of $\ell$ commuting
generators $h_{j_1}h_{j_2}\cdots h_{j_\ell}$ with $j_1<j_2<\cdots<j_\ell$, i.e.,
\be\label{charges}
H_M^{(0)} &=& \id\,,\non\\
H_M^{(1)} &=& -\mathcal{H} =
 \sum_{j=1}^Mh_j\,,\non\\
H_M^{(2)} &=& \sum_{j_1=1}^M
\sum_{j_2=j_1+p+1}^M h_{j_1}h_{j_2} \,,\\
&\vdots&\non\\
H_M^{({\overline M})} &=& \sum_{j_1=1}^M
\sum_{j_2=j_1+p+1}^M
\cdots \sum_{j_{\overline M}=
j_{{\overline M}-1}+p+1}^M h_{j_1}h_{j_2}\dots h_{j_{\overline M}}\non\,.
\ee

Associated to these charges we define the generating
function,
\be\label{transfer}
G_M(u) = \sum_{\ell=0}^{\overline M}(-u)^{\ell} H_M^{(\ell)}\,,
\ee
where $u \in \C$ is a spectral parameter. We will prove
that indeed 
\be\label{comcharges}
\left[H_M^{(\ell)},H_M^{(\ell')}\right] = 0\,\quad \forall \ell,\ell'\,,
\ee
and consequently generating functions with distinct spectral
parameters also commute, i.e.,
\be\label{comcharges}
\left[G(u),G(u')\right] = 0\,.
\ee

The key property to prove (\ref{comcharges})
follows from the fact that both, the charges (\ref{charges})
and the generating function
(\ref{transfer}), satisfy some recurrence relations that we now derive. 
Rewriting the last sum in $H_M^{(\ell)}$ in (\ref{charges}) as,
\begin{widetext}
\be
 \cdots \sum_{j_{l}=j_{{\ell}-1}+p+1}^M h_{j_1}h_{j_2}\dots h_{j_{\ell}}&=&
 \cdots \sum_{j_{\ell}=j_{{\ell}-1}+p+1}^{M-1} h_{j_1}h_{j_2}
\dots h_{j_{\ell}}+h_{j_1}\dots h_{j_{{\ell}-1}}h_M\,,
\ee
\end{widetext}
we obtain the recurrence relation,
\be\label{recurrencecharges}
H_M^{(\ell)} = H_{M-1}^{(\ell)}+h_M H_{M-(p+1)}^{(\ell-1)}\,,
\ee
with the initial conditions $H_M^{(0)}=\id$, $H_M^{(\ell)}=0$ for $\ell<0$ or $M\leq 0$.
It follows from (\ref{recurrencecharges}) and (\ref{transfer}) that
\be\label{grec1}
G_M(u) =
\sum_{\ell=0}^{\overline{M}} (-u)^{\ell} H_{M-1}^{(\ell)}
-uh_M\sum_{\ell=0}^{\overline{M} -1} (-u)^{\ell}
 H_{M-(p+1)}^{(\ell)},\non
\ee
where we have used $H_{M-(p+1)}^{(-1)} =0$.

Since, from (\ref{bar}) $\overline{M} -1 = \overline{M-(p+1)}$, we identify 
from (\ref{transfer}) the second summation  in (\ref{grec1}) as 
$G_{M-(p+1)}(u)$. Writing $M=j(p+1) +q$ with $j,q \in \Z$ and $0\leq q \leq p$,
we have that $\overline{M} = \overline{M-1}$ except for $q=1$ where 
$\overline{M} = \overline{M-1} +1$.
Since for $q=1$, $\overline{M}> \overline{M-1}$ we have that  $H_{M-1}^{(\overline{M})}=0$, 
and consequently we identify, for all $M$, the first summation in 
(\ref{grec1}) as $G_{M-1}(u)$. We have then the recurrence relation for the 
generating function
\be\label{recurrenceformula}
G_M(u) = G_{M-1}(u)-uh_M G_{M-(p+1)}(u)\,,
\ee
with $G_M(u)=\id$ for $M\leq 0$.

The recurrences (\ref{recurrencecharges}) and (\ref{recurrenceformula})
are the basic identities we use in this paper. It is convenient,
for further use, to iterate (\ref{recurrencecharges}) and (\ref{recurrenceformula}),
\be\label{it1}
G_{M-(p+1)}(u) &=& G_{M-(p+1)-j}(u) \\ &-&
u \sum_{k=0}^{j-1}h_{M-(p+1)-k}G_{M-2(p+1)-k}(u)\,,\non
\ee
\be\label{it3}
H_{M-(p+1)}^{(\ell)}&=&H_{M-(p+1)-j}^{(\ell)}\\&+&
\sum_{k=0}^{j-1}h_{M-(p+1)-k}H_{M-2(p+1)-k}^{(\ell-1)}\,,\non
\ee
for $j=0,1,2,\dots,M-(p+1)$, and also,
\be\label{it2}
G_{M-1}(u) = G_{M-(p+1)}(u) - u
\sum_{k=1}^{p}h_{M-k}G_{M-(p+1)-k}(u)\,,\non\\
\ee
\be\label{it4}
H_{M-1}^{(\ell)} = H_{M-(p+1)}^{(\ell)} +
\sum_{k=1}^p h_{M-k} H_{M-(p+1)-k}^{(\ell-1)}\,.
\ee

In order to demonstrate the involution (\ref{comcharges})
let us show initially that an arbitrary charge
commutes with the generating function. 
For this sake it is convenient to define,
\be\label{defbeta}
\beta_{M,q}^{(\ell)} \equiv \left[H_M^{(\ell)},G_M(u)\right]_q\,,\quad \ell=1,\dots,\overline M\,.
\ee
where
\be
\left[X,Y\right]_q = XY+q YX
\ee
is the $q$-commutator. We want to show that $\beta_{M,-}^{(\ell)} =0$, for 
any $\ell$.
In the remaining of this section, for simplicity, we will omit the explicit
dependence of the generating function in the variable $u$.

Inserting 
(\ref{recurrencecharges}) and (\ref{recurrenceformula}) in (\ref{defbeta})
with $q=-1$ we obtain,
\be\label{betaAB}
\beta_{M,-}^{(\ell)}=\beta_{M-1,-}^{(\ell)}-uh_M^2 \beta_{M-(p+1),-}^{(\ell-1)}
+ A + B
\ee
where
\be\label{auxA}
&&A = \left[h_MH_{M-(p+1)}^{(\ell-1)},G_{M-1}\right]_{-}\,,\\
&&B = -u\left[H_{M-1}^{(\ell)},h_MG_{M-(p+1)}\right]_{-}\,.
\ee
Using (\ref{it2}) in A and (\ref{it4}) in B, we obtain,
\be
A &=& h_M \beta_{M-(p+1),-}^{(\ell-1)} \\&-& u h_M
\sum_{j=1}^p
\left[H_{M-(p+1)}^{(\ell-1)},h_{M-j}G_{M-(p+1)-j}\right]_{-\omega}\non
\ee
and
\be
B&=&-uh_M\beta_{M-(p+1),-}^{(\ell)}\\&-&\omega uh_M
\sum_{j=1}^p
\left[h_{M-j}H_{M-(p+1)-j}^{(\ell-1)},G_{M-(p+1)}\right]_{-\omega^{-1}}\,.\non
\ee
Defining,
\bea
&&X_{M,q}^{(\ell)}(j) = \left[H_{M-(p+1)}^{(\ell)},h_{M-j}G_{M-(p+1)-j}\right]_{q}\,,\label{defX}\\ 
&&
\label{defY}
Y_{M,q}^{(\ell)}(j) = \left[h_{M-j}H_{M-(p+1)-j}^{(\ell)},G_{M-(p+1)}\right]_{q}\,,\\
&&
\label{defgamma}
\gamma_{M,q}^{(\ell)}(j) = X_{M,-q}^{(\ell)}(j)+qY_{M,-q^{-1}}^{(\ell)}(j)\,,
\eea
we can write,
\be
A+B &=& h_M \beta_{M-(p+1),-}^{(\ell-1)}-uh_M\beta_{M-(p+1),-}^{(\ell)}
\non\\&-&uh_M
\sum_{j=1}^p\gamma_{M,\omega}^{(\ell-1)}(j)\,.
\ee

We now obtain a recurrence relation for $X_{M,q}^{(\ell)}$,
$Y_{M,q}^{(\ell)}$ and $\gamma_{M,q}^{(\ell)}$. Inserting (\ref{it3}) in (\ref{defX})
and (\ref{it1}) in (\ref{defY}) we obtain, after straightforward manipulations,
\be\label{Xsimp}
X_{M,q}^{(\ell)}(j) &=&h_{M-j}\beta_{M-(p+1)-j,q}^{(\ell)}
\non\\&+&\omega
h_{M-j}
\sum_{k=0}^{j-1}Y_{M-j,q\omega^{-1}}^{(\ell-1)}(p+1+k-j)\,,
\ee
\be\label{Ysimp}
Y_{M,q}^{(\ell)}(j) &=&h_{M-j}\beta_{M-(p+1)-j,q}^{(\ell)}
\non\\&-&uh_{M-j}
\sum_{k=0}^{j-1}X_{M-j,q\omega}^{(\ell)}(p+1+k-j)\,,
\ee
Using (\ref{Ysimp}) in (\ref{Xsimp}) and (\ref{Xsimp})
in (\ref{Ysimp}), with $q=\omega$, we obtain
\bea\label{gammasimp}
&&\gamma_{M,\omega}^{(\ell)}(j) = 
 (1+q)h_{M-j}\beta_{M-(p+1)-j,-}^{(\ell)} \nonumber \\
&& +
\omega h_{M-j} 
 \sum_{k=0}^{j-1}h_{M-(p+1)-k}
\left(\beta_{M-2(p+1)-k,-}^{(\ell-1)} \right.\nonumber \\
&&- 
\left. u\beta_{M-2(p+1)-k,-}^{(\ell)} \right. \nonumber \\
&& \left. -u\sum_{k'=0}^{p+k-j}\gamma_{M-(p+1)-k,\omega}^{(\ell-1)}(k'-k+j)\right)\,.
\eea
We then finally obtain from (\ref{defgamma}) and (\ref{betaAB}),
\be\label{betasimp1}
\beta_{M,-}^{(\ell)} &=& \beta_{M-1,-}^{(\ell)}-uh_M\beta_{M-(p+1),-}^{(\ell)}
-uh_M\sum_{j=1}^p \gamma_{M,\omega}^{(\ell-1)}(j)
\non\\&-&h_M\beta_{M-(p+1),-}^{(\ell-1)}(uh_M-1)
\,,
\ee
with $\gamma_{M,\omega}^{(\ell)}(j)$ given by (\ref{gammasimp}).

We have from
(\ref{charges}) that
$H_M^{(l)}=\beta_{M,-}^{(l)}=\left[H_M^{(l)},G_M\right]=0$ for $l\leq 0$,
and $G_M^{(l)}=1$ for $l \leq 0$, therefore from (\ref{gammasimp})
$\gamma_{M,q}^{(l)}=0$ for $l\leq 0$.

Inserting these values in (\ref{betasimp1}) we obtain for $l=1$\,,
\be\label{betarec1}
\beta_{M,-}^{(1)} = \beta_{M-1,-}^{(1)}-uh_M\beta_{M-(p+1),-}^{(1)}\,.
\ee
Since for $M=1$, $\beta_{l,-}^{(1)}=0 $ ($l\leq 0$), we obtain $\beta_{1,-}^{(1)}=0$
and by iterating (\ref{betarec1}) we obtain
$\beta_2^{(1)}=\beta_3^{(1)}=\cdots=\beta_M^{(1)}=0$,
for arbitrary $M$. From (\ref{gammasimp}) $\gamma_{M,\omega}^{(1)}(j)$
only depends on $\beta_{M,-}^{(0)}$,
$\beta_{M,-}^{(1)}$ and $\gamma_{M,\omega}^{(0)}$.
This means that $\gamma_{M,\omega}^{(1)}(j)=0$
for all $M$ and from (\ref{betasimp1}) with $l=2$ we have,
\be
\beta_{M,-}^{(2)} = \beta_{M-1,-}^{(2)}-uh_M \beta_{M-(p+1),-}^{(2)}\,.
\ee
Since $\beta_{l,-}^{(l)}=0$ for
$l\leq 0$ we have $\beta_{1,-}^{(2)}=\beta_{2,-}^{(2)}=\cdots=\beta_{M,-}^{(2)}=0$,
for all $M$. Similarly as before $\gamma_{M,\omega}^{(2)}$
only depends on products involving 
$\beta_{M,-}^{(1)}$, $\beta_{M,-}^{(2)}$ and
$\gamma_{M,\omega}^{(1)}$ therefore $\gamma_{M,\omega}^{(2)}=0$
and then from (\ref{gammasimp}),
\be
\beta_{M,-}^{(3)} = \beta_{M-1,-}^{(3)}-uh_M \beta_{M-(p+1),-}^{(3)}\,.
\ee
Since $\beta_{\ell,-}^{(3)}$ for $\ell\leq 0$ we have $\beta_{M,-}^{(3)}$ for all $M$.
This procedure iterates
and we have our proof:
\be\label{commuteHT}
\beta_{M,-}^{(\ell)}=\left[H_M^{(\ell)},G_M(u)\right]=0\,\quad \forall 
\ell=1,\dots,M.
\ee
Expanding this result in powers of $u$ we obtain that all the distinct charges
commute among themselves, \textit{i.e.},
\be\label{commuteHH}
\left[H_M^{(\ell)},H_M^{(\ell')}\right] = 0\,\quad \forall \ell,\ell'\,.
\ee
The relation (\ref{commuteHH}) also imply that the generating function
with arbitrary values of the spectral parameter $u$ commute, \textit{i. e.},
\be\label{commuteTT}
\left[G_M(u),G_M(v)\right] = 0\,.
\ee
The relations (\ref{commuteHH}) or (\ref{commuteTT}) imply that
the Hamiltonian (\ref{Hgen}) with generators $\{h_i\}$ satisfying (\ref{halgebra1})
is exactly integrable. {\it The closure relation (\ref{halgebra3}) is not a necessary
condition for ensuring the exact integrability}.

\section{The inversion relation and product formula for the generating function}\label{sec:prod}

In this Section, we show that when the generators
$\{h_i\}$ defining the Hamiltonian (\ref{Hgen}) satisfy
the closure relation (\ref{halgebra3}) besides the relations
(\ref{halgebra1}), the
generating function (\ref{transfer}) satisfies,
\be\label{prodaux}
G_M(u)G_M(\omega u)\cdots G_M(\omega^{N-1} u) =
P_M^{(p)}(u^N)\id
\ee
where $P_M^{(p)}(u^N)$ is a polynomial of degree $\overline M$
in $u^N$. Equation (\ref{prodaux}) implies (see Sec. \ref{sec:pol}), that
the Hamiltonians
have a free fermion ($N=2$)
 or a free parafermionic eigenspectrum ($N>2$).

\subsection{ N=2}

Let us first consider the free fermion case ($N=2$). This case
has been considered for $p=1$ in the context
of the $\tau_2$ model \cite{Baxfunc,Baxter2014} and for $p=2$
in \cite{Fendley:2019} using a certain factorization of the generating
function. Here we prove it for any
value of $p\geq 1$ by showing that
\be\label{tau}
\tau_M^{(2)}(u)\equiv G_M(u)G_M(-u)= P_M^{(p)}(u^2)\id\,,
\ee
satisfies a recurrence relation.

In order to
simplify the notation let us define the algebraic operation,
\be
\mathcal{L}(A(u);B(u)) = A(u)B(-u)-B(u)A(-u)\,.
\ee
Using the fundamental relation (\ref{recurrenceformula}) in (\ref{tau})
we obtain,
\be
\tau_M^{(2)}(u) = \tau_{M-1}^{(2)}(u)-u^2h_M^2
\tau_{M-(p+1)}^{(2)}(u)+\Xi_M^{(2)}(u)\,,
\ee
where
\be\label{defXi}
\Xi_M^{(2)}(u) = -uh_M\sum_{j=1}^p L_{M,j}(u)\,,
\ee
with
\be\label{defL}
L_{M,j}(u) \equiv \mathcal{L}(h_{M-(p+1)-j}G_{M-2(p+1)+j}(u);G_{M-(p+1)}(u)).\non\\
\ee
Using (\ref{it1}) for $G_{M-(p+1)}$ in (\ref{defL}) we obtain for $j=1,\dots,p$,
\be\label{recL}
L_{M,j}(u) &=& L_{M-1,j+1}(u)\non\\&-&uh_{M-(p+1)+j}L_{M-(p+1)-j,p+1-j}(u)\,,
\ee
with the condition $L_{M,p+1}(u)=0$ for all $M$.
Since $G_M(u)=1$ and $h_M=0$ for $M\leq 0$, we have from (\ref{defL}),
\be
L_{M,j}(u)=0\quad \textrm{for} \quad M\leq p+1,\quad j=1,\dots,p\,.
\ee
The recurrence relation (\ref{recL}) then imply,
\be
L_{M,j}(u)=\Xi_M(u) = 0\quad \forall M\,,
\ee
and
\be\label{rectau}
\tau_M^{(2)}(u)=\tau_{M-1}^{(2)}(u)-u^2h_M^2\tau_{M-(p+1)}^{(2)}(u)\,,
\ee
with $\tau_M^{(2)}(u)=1$ for $M\leq 0$. This last expression
gives $\tau_1^{(2)}(u)=1-u^2h_1^2$ and by iterating (\ref{rectau})
we obtain that $\tau_M^{(2)}(u)$ is a polynomial in $u^2$,
reproducing (\ref{tau}) for arbitrary values of $p$.

As a consequence of the recurrence  (\ref{rectau}) the polynomial
$P_M^{(p)}(u^2)$ also satisfy the recurrence,
\be\label{recpol}
P_M^{(p)}(u^2)=P_{M-1}^{(p)}(u^2)-u^2\lambda_M^2P_{M-(p+1)}^{(p)}(u^2)\,,
\ee
with the initial condition $P_M^{(p)}(u^2)=1$ for $M\leq 0$. 

It is important
to notice from (\ref{recpol}) that the polynomials
$P_M^{(p)}(z)=P_M^{(p)}(\{\lambda_i^2\},z)$ depend
on the parameters $\{\lambda_i^2\}$ defining the closure relation
(\ref{halgebra3}) and the coupling constants of the
Hamiltonian
(\ref{Hgen}).

\subsection{  N=3}

We now consider the case $N=3$. Extending the definition (\ref{tau}) for the 
case $N=2$ we now define
\be\label{tau3}
\tau_M^{(3)}(u) \equiv G_M(u)G_M(\omega u)G_M(\omega^2 u) = P_M^{(p)}(u^3)\id
\ee
where $\omega=e^{2i\pi/3}$ and, as we shall see, $P_M^{(p)}(u^3)$ is related to the same
polynomial of degree $\overline M$ appearing in (\ref{tau}) for the $Z(2)$ case.

In Appendix \ref{appendixA}, we show the recurrence relation
\be\label{rectau3}
\tau_M^{(3)}(u)=\tau_{M-1}^{(3)}(u)-u^3h_M^3\tau_{M-(p+1)}^{(3)}(u)\,.
\ee
Since $\tau_1^{(3)}(u)=\id$, we obtain that $\tau_M^{(3)}(u)$ is given by the 
polynomial $P_M^{(p)}(u^3)$ as claimed in (\ref{tau3}). Also from (\ref{halgebra3})
$h_M^3=\lambda_M^3$ the polynomial satisfies
the recurrence relation
\be\label{recpol3}
P_M^{(p)}(u^3)=P_{M-1}^{(p)}(u^3)-u^3\lambda_M^3P_{M-(p+1)}^{(p)}(u^3)\,,
\ee
with $P_{M'}^{(p)}=1$ for $M' \leq 0$. We can see that the polynomials $P_M^{(p)}(z)=P_M^{(p)}(\{\lambda_i^3\},z)$
are the same ones that appeared in the case $N=2$, where we replace the couplings
$\{\lambda_i^2\}$ by $\{\lambda_i^3\}$.

\subsection{  N$>$3 }

For $N>3$, we can proceed in a similar way, and show that,
\be\label{prodN}
\tau_M^{(N)}(u) = G_M(u)G_M(\omega u)\cdots G_M(\omega^{N-1} u) =
P_M^{(p)}(u^N)\id\non\\
\ee
with $\omega=e^{2i\pi/N}$, satisfies the recurrence relation,
\be\label{rectauN}
\tau_M^{(N)}(u)=\tau_{M-1}^{(N)}(u)-u^Nh_M^N\tau_{M-(p+1)}^{(N)}(u)\,,
\ee
with $\tau_M^{(N)}(u)=1$ for $M\leq 0$. From (\ref{rectauN})
we obtain the recurrence relation for the polynomials
$P_M^{(p)}(u^N)=P_M^{(p)}(\{\lambda_i^N\},u^N)$,
\be\label{recpolN}
P_M^{(p)}(u^N)=P_{M-1}^{(p)}(u^N)-u^N\lambda_M^NP_{M-(p+1)}^{(p)}(u^N)\,,
\ee
with $P_{M'}^{(p)} =1$ for $M' \leq 0$.
A proof of (\ref{rectauN}), similar as we did for the cases $N=2$
and $N=3$ (see Appendix \ref{appendixA}), for general $N$, is straightforward but lengthy. Anyway, we
have checked (\ref{rectauN}) and hence (\ref{prodN})
for several values of $N$ and lattice sizes $M$.

Comparing the recurrence relations (\ref{rectauN}) and (\ref{recpolN}) we identify
the coefficients $C_M^{(l,p)}$ in the expansion,
\be\label{expP}
P_M^{(p)}(z)=\sum_{\ell=0}^{\overline M}(-z)^\ell C_M^{(\ell,p)}\,.
\ee
by replacing $h_j\leftrightarrow \lambda_j^N$ in (\ref{charges}), i.e.,
\be\label{polcoef}
C_M^{(\ell,p)} = \sum_{j_1=1}^M\sum_{j_2=j_1+p+1}^M
\cdots \sum_{j_{\ell}=j_{{\ell}-1}+p+1}^M \lambda_{j_1}^N\lambda_{j_2}^N\dots \lambda_{j_\ell}^N \,,
\ee
for $\ell=0,1,\dots,\overline M$.

In the case where all $\lambda_i^N=1$, $C_M^{(\ell)}$
is the number of distinct ways we can put $\ell$ particles with excluded
volume of $(p+1)$ lattice units in a lattice with $M$ sites, i.e.,
\be
C_M^{(\ell,p)}=\binom{M-p(\ell-1)}{\ell}=\frac{(M-p(\ell-1))!}{(M-p(\ell-1)-\ell)!\ell!}\,,
\ee
and $P_M^{(p)}(z)$ is the generalized hypergeometric polynomial
known as ${}_{p+1}F_p$ \cite{AeqB}:
\begin{widetext}
\be\label{polP}
P_M^{(p)}(z) = {}_{p+1}F_p \left(\begin{matrix} -\frac{M+p}{p+1}
\quad -\frac{M+p-1}{p+1} \quad -\frac{M+p-2}{p+1} \quad \cdots \quad -\frac{M}{p+1}\\
-\frac{M+p}{p} \quad -\frac{M+p-1}{p} \quad \cdots \quad -\frac{M+1}{p}\end{matrix}\,;
 \frac{(p+1)^{p+1}}{p^p}z\right)
= \sum_{\ell=0}^{\overline M}(-1)^\ell\binom{M-p(1-\ell)}{\ell}z^{\ell}\,.
\ee
\end{widetext}

In this symmetric case the polynomial $P_M^{(p)}(-\tilde z)$, as 
we can see from (\ref{expP}), is the grand canonical partition
function of a polymer with monomers with size of $(p+1)$ lattice units
and fugacity $\tilde z$ in a lattice of $M$ sites. As we are going to see
in the following sections the roots of $P_M^{(p)}(-\tilde z)$ for
any $p$ and finite $M$ are real and negative.
This
means that the grand canonical partition function is analytic. However
as $M\rightarrow\infty$ the largest root approaches 0 and therefore
in the thermodynamic limit the polymer has a critical fugacity $\tilde z=
\tilde{z}_c=0$.

\section{The eigenspectrum of the free-fermionic and free-parafermionic quantum chains}\label{sec:pol}

The eigenspectrum of any quantum chain (\ref{Hgen}) expressed in terms of the 
generators $\{h_i\}$ satisfying (\ref{halgebra1})-(\ref{halgebra3}) are obtained from the zeros 
of the fundamental polynomials 
$P_M^{(p)}(z) \equiv P_M^{(p)} (\{\lambda_i\},u^N)$ derived in the last section.
This means that for a given parameter $p$, all the models with arbitrary $N$ 
are ruled by the same polynomial. 

It is interesting to mention that in the case where $p=1$ and $\lambda_i=1$ 
($i=1,\ldots,M$), that includes the critical quantum Ising chain (\ref{HIsing}) and 
the critical $Z(N)$ free-parafermionic Baxter chain (\ref{HBaxter}), these polynomials 
are related to the well known Chebyshev polynomial of second type, \textit{i.e.},
$P_M^{(1)}(z) = z^{\frac{M+1}{2}}U_{M+1}\left(\frac{1}{2z^\frac{1}{2}}\right)$.
In oder to illustrate we present in Table \ref{tpol1} some polynomials at 
$\{\lambda_i=1\}$ in the cases $p=1,2,3$.

\begin{table*}[t] 
\centering
\begin{tabular}{|l|l|l|l|}
\hline \hline
 $M$ & $P_M^{(1)}(z)$& $P_M^{(2)}(z)$ & $P_M^{(3)}$  \\ \hline 
1 & $1-z$ & $1-z$ & $1-z$ \\
 2 & $1-2 z$ & $1-2z$ & $1-2z$ \\
 3 & $1-3 z+z^2$ & $1-3z$ & $1-3z$ \\
 4 & $1-4 z+3 z^2$ & $1-4z + z^2$ &$1-4z$\\
 5 & $1-5 z+6 z^2-z^3$ & $1-5z +3z^2$ & $1-5z +z^2$\\
 6 & $1-6 z+10 z^2-4 z^3$ & $1-6z +6z^2$ & $1- 6z+6z^2$\\
 7 & $1-7 z+15 z^2-10 z^3+z^4$ & $1-7z +10z^2 -z^3$ & $1-7z+6z^2$ \\
 8 & $1-8 z+21 z^2-20 z^3+5 z^4$ & $1-8z+15z^2 -4z^3$ & $1-8z+10z^2$ \\
 9 & $1-9 z+28 z^2-35 z^3+15 z^4-z^5$ & $1-9z+21z^2-10z^3$ & $1-9z+15z^2 -z^3$ 
\\
10 & $1-10 z+36 z^2-56 z^3+35 z^4-6 z^5$ & $1-10z+28z^2-20z^3+z^4$ &
$1-10z-21z^2-4z^3$  \\
 11 & $1-11 z+45 z^2-84 z^3+70 z^4-21 z^5+z^6$ & $1-11 z+36 z^2-35 z^3+5 z^4$ &
 $1 -11z +28z^2 -10z^3$  \\\hline
\end{tabular}
\caption{Example of polymials $P_M^{(p)}(z)$i, with coupling constants 
$\{\lambda_i=1\}$, for $p=1,2$ and $p=3$, and for $M=1-11$.}
\label{tpol1}
\end{table*}



Since $[G_M(u),G_M(-u)]=0$ and $G_M(0)=P_M^{(p)}(0)=1$, by applying in the product formula
(\ref{prodaux}) to a 
given eigenfunction of $G_M(u)$, with eigenvalue $\Lambda_M(u)$, we obtain
\be
\label{prodlam}
\Lambda(u)\dots\Lambda(\omega^{N-1}u)=P_M^{(p)}(u^N)=\prod_{i=1}^{\overline M}\left(1-\frac{u^N}{z_i}\right)\,,
\ee
where $z_i$ are the roots of $P_M^{(p)}(z_i)=0$. Solving (\ref{prodlam}) in terms 
of $z_i$ we obtain,
\be\label{Lam}
\Lambda_M(u) = \prod_{i=1}^{\overline M} \left(1- u\frac{\omega^{s_i}}{z_i^{1/N}}\right)
= \prod_{i=1}^{\overline M} \left(1- u\,\omega^{s_i}\epsilon_i\right)\,,
\ee
where $\epsilon_i=z_i^{-1/N}$ and $s_i \in \{0,1,\dots,N-1\}$. In all the cases we
considered in this paper, where the couplings are real, we verified that the roots $z_i$ are real 
positive  and distinct,
implying the 
existence of $N^{\overline{M}}$ distinct eigenvalues for the generating function 
$G_M(u)$. We can also expand (\ref{Lam}) in powers of $u$ \cite{Vinberg}, 
\be\label{lamsol}
\Lambda_M^{\{s_i\}}(u) =
\sum_{\ell=0}^{\overline M} (-1)^\ell e_\ell\left(\omega^{s_1}\epsilon_1,\dots,\omega^{s_{\overline M}}\epsilon_{\overline M}\right)u^\ell\,,
\ee
where 
\be \label{5.4}
e_{\ell}(x_1,x_2,\dots,x_n)=\sum_{1\leq j_1 < j_2 < \cdots < j_\ell \leq n }  x_{j_1}x_{j_2}\cdots x_{j_\ell}\non\\
\ee
for $\ell=0,1,\dots,n$ is the $\ell$-th elementary symmetric polynomial in the variables 
$x_1,x_2,\ldots,x_n$.

Applying an eigenfunction of $G_M(u)$ with eigenvalue $\Lambda_M^{\{s_i\}}(u)$ in
(\ref{transfer}), since $u$ is arbitrary, (\ref{lamsol}) implies that the eigenfunction is 
also an eigenfunction of the charges $\{H_M^{(\ell)}\}$ given in (\ref{charges}) with 
eigenvalues 
\be \label{othercharges}
q_{\{s_i\}}^{(\ell)}=
e_l\left(\omega^{s_1}\epsilon_1,\dots,\omega^{s_{\overline M}}\epsilon_{\overline M}\right),\quad
\ell=1,\dots,\overline M\,.
\ee
In particular the Hamiltonian (\ref{Hgen}) has the free-fermionic ($N=2$) of 
free-parafermionic ($N>2$) spectrum, with eigenenergies
\beq\label{EHgen}
-E^{\{s_i\}} = q_{\{s_i\}}^{(1)}= \omega^{s_1} \epsilon_1 +
 \omega^{s_2} \epsilon_2 + \cdots + \omega^{s_L} \epsilon_{\overline M}\,.
\eeq
This means that not only  the eigenspectrum of the general Hamiltonians (\ref{Hgen}), but 
also the the ones of the extended Hamiltonians $\{H_M^{(\ell)}\}$, are given entirely in terms 
of the roots of the fundamental polynomials $P_M^{(p)}(\{\lambda_i\},z)$. 

\begin{figure} [htb]
\centering
\includegraphics[width=0.45\textwidth]{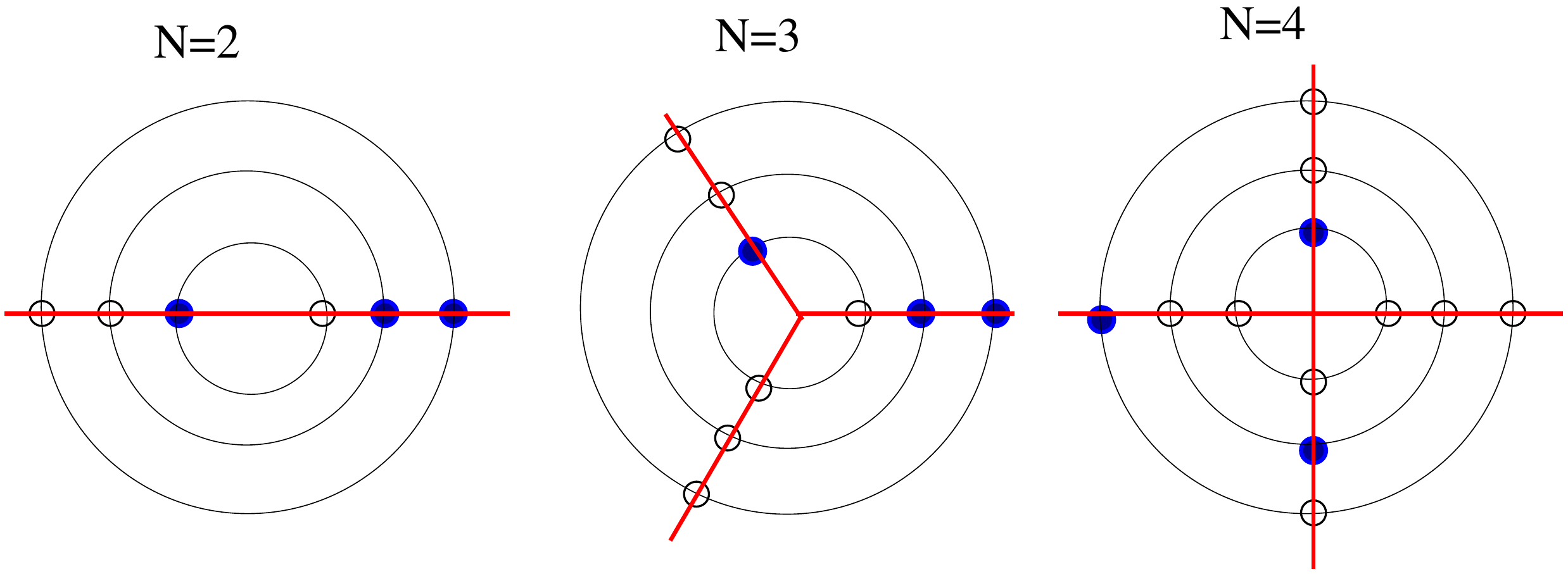}
\caption{  
Schematic representation of the eigenenergies of a quantum
chain with $\overline{M}=3$ and $N=2,3,4$. 
There are 3 quasi-energies, that fixes 
the radius of the $\overline{M}$ circles in the complex plane. There is a 
"circle exclusion principle"  that imposes a single quasi-energy in each circle 
(filled circles). The figures show for $N=2$ and $N=3$
the ground-state and first excited state energies, respectively.} \label{fig1}
\end{figure}

In Fig.~\ref{fig1} we show schematically examples of  eigenenergies for a 
$Z(N)$ quantum
chain with $\overline{M}=3$ and $N=2,3,4$. 
In Appendix B  we consider, as an example, the simple case of the $Z(N)$ Hamiltonian  
with $p=3$ and with $M=5$.

In the next section, by exploiting the solution for the polynomial roots of
$P_M^{(p)}(z)$ we are going to derive  the critical behavior of the Hamiltonian
(\ref{Hgen}) at a special critical point.

\section{The ground state energy and critical exponents for the quantum chains}\label{sec:crit}

For general values of the couplings $\{\lambda_i\}$ we should expect a quite 
rich phase diagram for the Hamiltonians (\ref{Hgen}) with $p\geq2$.

We restrict ourselves  to  the quantum chains at their symmetrical point where all the 
couplings constants  $\lambda_i=1$, ($i=1,\ldots,M$) in (\ref{halgebra3}) or any of its 
representations like (\ref{HBaxter},\ref{HP},\ref{HA}).

For $p=1$, where for $N=2$ the possible representations are the free 
fermionic Ising quantum chain (\ref{HIsing}) and (\ref{HIsingeven}), and for $N>2$ the 
$Z(N)$ free parafermionc quantum chain (\ref{HBaxter}), the models are critical 
with a dynamical critical exponent $z=2/N$ and specific-heat exponent 
$\alpha = 1-2/N$ \cite{Alcaraz_2017}. In \cite{Fendley:2019} it was shown
that in the particular case $p=N=2$ the symmetrical point is a 
multicritical point where $z=3/2$.

In \cite{AP2020} we showed that for 
general values of $p$ the polynomial roots $\{z_i\}$ of $P_M^{(p)}$, that give 
us the quasi-particles energies $\epsilon_i=z_i^{-1/N}$  in (\ref{EHgen}), for 
arbitrary M, and lattice sizes multiples of $(p+1)$, can be parametrized by 
trigonometric functions
\beq\label{epansatz}
\epsilon_k =\frac{\sin^{\frac{p+1}{N}}(\mathfrak{p}_k)}{\sin^{\frac{1}{N}}
\left(\frac{\mathfrak{p}_k}{p+1}\right) \sin^{\frac{p}{N}}
\left(\frac{p\mathfrak{p}_k}{p+1}\right)}\,,\quad k=1,\dots,\overline M\, ,
\eeq
where $\mathfrak{p}_k$ is a quantum number. For $p=1$,
$\mathfrak{p}_k=k\pi/(\overline M + 1)$, for any $M$, and for $p>1$ we conjectured,
and confirmed numerically,
that as $M\to \infty$ the distribution density for the quantum  numbers behaves 
as $\Delta \mathfrak{p}_k/\Delta k = \pi/{\overline M} = (p+1)\pi/M$. Since the 
ground state energy is obtained by taking in (\ref{EHgen}) the values $s_i=0$ 
($i=1,\ldots,\overline M$), we obtain an exact expression for the ground-state 
energy per site:
\beq\label{einf}
e_{\infty}^{(p)} \equiv -\frac{E_0}{M}=
 -\frac{1}{M}\sum_{k=1}^{\overline M}\epsilon_k=-\frac{1}{(p+1)\pi}
\int_{0}^\pi\epsilon(\mathfrak{p})d\mathfrak{p}\,.
\eeq
For $N=2$ and general values of $p$, (\ref{einf}) is given in terms of gamma 
functions:
\beq\label{einfn2}
e_{\infty}^{(p)} =
 -\frac{\Gamma\left(\frac{1}{2}+\frac{p}{2}\right)}{\Gamma\left(1+\frac{p}{2}\right)}\,\quad (N=2).
\eeq
For general $p$ and $N$ \cite{AP2020}, the integral 
(\ref{einf}) is expressed in 
terms of integral representation of the Lauricella hypergeometric 
series $F_D^{(p-1)}$\cite{Slater}. Moreover, in the cases $p=1,2$ and 3 
(\ref{einf}) is given in terms of the gamma, ${}_2F_1$ and Appel functions $F_1$, i. e.,
\beq\label{ep1}
e_{\infty}^{(1)} = -\frac{2^{\frac{2}{N}-1}
\Gamma\left(\frac{1}{N}+\frac{1}{2}\right)}{\sqrt{\pi}\left(\frac{1}{N}+1\right)}\,,
\eeq
\beq\label{ep2}
e_{\infty}^{(2)} = -\frac{3^{\frac{3}{N}+\frac{1}{2}}\Gamma\left(\frac{3}{N}+1\right)}
{2^{\frac{2}{N}+2}\sqrt{\pi}\Gamma\left(\frac{3}{N}+\frac{3}{2}\right)}
{}_2F_1\left(\begin{matrix}\frac{1}{2}\quad\frac{1}{N}+\frac{1}{2}\\
 \frac{3}{N}+\frac{3}{2}\end{matrix} ;\frac{3}{4}\right)\,,
\eeq
\be\label{ep3}
e_{\infty}^{(3)} &=& -\frac{2^{\frac{8}{N}-\frac{3}{2}}\Gamma\left(\frac{4}{N}+1\right)}
{3^{\frac{3}{N}}\sqrt{\pi}\Gamma\left(\frac{4}{N}+\frac{3}{2}\right)}
\non\\&\times&
F_1\left(\frac{1}{2};\frac{1}{2}-
\frac{2}{N},\frac{3}{N};\frac{4}{N}+\frac{3}{2};\frac{1}{2},\frac{2}{3}\right)\; .
\ee
In the case $p=4$, we have 
\be\label{ep4}
&&e_{\infty}^{(4)} =
 -\frac{5^{\frac{5}{N}}\sin\left(\frac{\pi}{5}\right)\Gamma\left(\frac{5}{N}+1\right)}
{2^{\frac{8}{N}+1}\sqrt{\pi}\Gamma\left(\frac{5}{N}+\frac{3}{2}\right)}
\non\\&&\times
F_D^{(3)}\left(\frac{1}{2};
\frac{1}{2}+\frac{2}{N},-\frac{5}{N},\frac{4}{N};\frac{5}{N}+\frac{3}{2};
x_1,x_2,x_3\right)\,
\ee
where $F_D^{(3)}$ is the Lauricella function with 3
 variables at $x_1=\frac{1}{2+\frac{2}{\sqrt{5}}}$,
$x_2=\frac{2}{3+\sqrt{5}}$ and $x_3=\frac{1}{1+\frac{1}{\sqrt{5}}}$. For a 
comparison of (\ref{einfn2})-(\ref{ep4}) with the numerical results obtained from the
direct solution of the polynomial zeros $\{z_i\}$ we refer to 
\cite{AP2020}.

The dynamical critical exponent is evaluated from the finite-size behavior of 
the  mass gaps of the quantum Hamiltonians. The first excited energy state is 
obtained in (\ref{EHgen}) by taking the set $s_1=s_2=\cdots=s_{\overline{M}-1}=0$ 
and $s_{\overline{M}} = 1$. However  $\epsilon_{\overline{M}}$ is the smallest 
quasienergy, whose associated quantum number, for $M \to \infty$, behaves as 
$\mathfrak{p}_{\overline{M}} = \pi -a/M$, where $a$ is a harmless constant. 
threfore the  
real part of the energy gap (complex for $N>2$) has the leading behavior,
\beq\label{gap}
\Delta_M^{(p)}=\mbox{Re}(E_1-E_0)=
\mbox{Re}(1-\omega)\epsilon(\mathfrak{p}_{\overline M}) 
\approx
\left(\frac{a}{M}\right)^{z}\,,
\eeq
where 
\beq \label{6.9}
z=(p+1)/N 
\eeq
is the dynamical critical exponent.

For the case $p=1$  (\ref{6.9}) recovers for $N=2$  the known result for the conformally 
invariant quantum Ising chain ($z=1$) and for the $Z(N)$, the  Baxter 
free-parafermionic model ($z=2/N$), as calculated in \cite{Alcaraz_2017}. 
The case $p=N=2$ recovers the result $z=3/2$ derived in \cite{Fendley:2019}. 
In summary, all the free interacting quantum chains (\ref{Hgen}) at their 
symmetric point $\{\lambda_i=1\}$ are critical with the dynamical critical 
exponent given by (\ref{6.9}). 
Since $z$ is an increasing function of $p$ the correlation length of the critical chains,
that goes as $M^z$, increases for a given lattice size $M$, as we increase 
the parameter $p$. This is physically expected since we increase the range of 
non commuting operators in the Hamiltonian, and hence the quantum correlations.

In order to better characterize the critical universality classes at this critical 
point $\{\lambda_i=1\}$ for the general free interacting models (\ref{Hgen}) let us 
perturb the couplings $\{\lambda_i\}$ around $\{\lambda_i=1\}$. 

To simplify 
let us restrict our analysis for the cases where $M$ is a multiple of $(p+1)$,
\textit{i.e.}, $M=(p+1)\overline{M}$.
We consider perturbed Hamiltonians where all the couplings are kept at 
$\lambda_i=1$, except for the couplings $\lambda_{(p+1)k}=\lambda$, with
$k=1,\ldots$. For example for $p=1$ the sequence of couplings are 
($1,\lambda,1,\lambda,\ldots,1,\lambda$) and for $p=2$ we have 
($1,1,\lambda,1,1,\lambda,\ldots,1,1,\lambda$).

In general, for anisotropic scaling, the dynamical critical exponent is given by
$z=\nu_{\perp}/\nu_{\parallel}$, where $\nu_{\perp}$ and $\nu_{\parallel}$
are the correlation 
length exponents in the time and space directions, respectively.

The specific heat at the critical point, from the finite-size scaling theory (FSS)
of critical behavior \cite{Barber}, should have, as $M \to \infty$, the power-law behavior
\be \label{6.10}
C_M(\lambda=1) \approx M^{\alpha/{\nu_{\parallel}}},
\ee
specified by the critical exponent $\alpha$. The specific heat is given by 
the second derivative of the ground-state energy \cite{BH} 
\be \label{6.11}
C(\lambda,M) = -\frac{\lambda^2}{M} \frac{\partial^2 E_0(\lambda,M)}
{\partial \lambda^2} = 
-\frac{\lambda^2}{M} \sum_{i=1}^{\overline{M}} 
\frac{\partial^2 \epsilon_i}{\partial \lambda^2}.
\ee
Solving for the zeroes $\{z_i\}$ of the polynomials $P_M^{(p)}(z)$ we obtain 
the quasi energies $\epsilon_i(\lambda,M)$. 

Before considering the cases $p>1$, let us consider the case $p=1$. In this case 
we verified {\it surprisingly} that the corresponding generalized Chebyshev 
polynomials have exact zeros $z_i$, producing quasienergies,
\be\label{6.12} 
\epsilon_j=z_j^{-1/N},\quad z_j^{-1}= 1 +\lambda^N +2\lambda^{N/2}\cos{k_j},
\ee
where $k_j=2\pi j/(M+2)$, for finite $M$ and arbitrary $\lambda$. We recall that
for $N=2$ these are the quasienergies for the Hamiltonian (\ref{HIsingeven})
describing an Ising quantum chain with an impurity at one of its ends. In the 
case where $M$ is odd, whose representation (\ref{HIsing}) is the standard 
Ising quantum chain, the roots $\{z_i\}$ are exactly known only at $\lambda=1$ 
\cite{Alcaraz_2017}.

The specific heat is obtained from (\ref{6.11}) and (\ref{6.12}):
\begin{widetext}
\be \label{6.12}
C(\lambda,M) = -\frac{\lambda^2}{M} \sum_{j=1}^{\overline{M}} 
\left\{ (1-N) \frac{(\lambda^{N-1} + \lambda^{\frac{N}{2}-1} \cos k_j)^2}
{(1+\lambda^N + 2 \lambda^{N/2} \cos k_j)^{2-1/N}} \right. 
&+& \left. \frac{(N-1)\lambda^{N/2} + (N/2-1)\lambda^{N/2-2}\cos k_j}
{(1 + \lambda^N + 2\lambda^{N/2} \cos k_j)^{1-1/N}} \right\}.
\ee
\end{widetext}
At $\lambda=1$, since $k_j=2\pi j/(M+2)$, all terms in the above sums are of 
$o(1)$, except for the ones where $j=\overline{M} -k = M/2 -k$ with $k$ of $o(1)$. 
These last terms will dominate the sum. Since for these terms 
$1 + \cos k_j \sim o(1/M^2)$, we get for $M \to \infty$
\be \label{6.13}
C(1,M) \sim \frac{1}{M} (M^2)^{1-1/N} \sim M^{1-2/N},
\ee
giving us the critical exponent
\beq \label{6.14}
\alpha= 1-2/N.
\eeq

This exponent can also be derived for the case where $M$ is odd, but the lack 
of an exact expression for the zeroes, renders the derivation  lenghty.

In the general cases $p>1$ we do not have an analytical solution for the 
roots as in (\ref{6.12}) and we have to evaluate them numerically. We also have 
to calculate numerically the derivatives of the quasi-particle energies 
$\epsilon_i$ at $\lambda=1$. By taking the specific heat values at two distinct 
lattice sizes $M_1$ and $M_2$ we produce the finite-size estimator:
\beq \label{6.15}
\alpha_{M_1,M_2} = \frac{\ln(C(1,M_1)/C(1,M_2))}{\ln(M_1/M_2)},
\eeq
that should tend towards $\alpha$ as $M_1,M_2 \to \infty$. In Table \ref{table1}
we give the results obtained by extrapolating sequences of $\alpha_{M_1,M_2}$ 
for the models with $p=2,\dots,5$ and $N=2,\dots,9$. The results were obtained by using 
van den Broeck Schwartz \cite{vbs} extrapolants up to lattice sizes  
$M_{\scriptsize{\mbox{max}}}$, shown in the last line of the table.
The zeroes of $P_M^{(p)}(z)$ were numerically evaluated with 50 decimal 
digits by using multiple precision calculations.

Our results indicate the conjecture
\beq \label{6.16} 
\alpha = \max{\{0,1-(p+1)/N\}},
\eeq
\textit{i.e.}, the models will have a vanishing critical exponent for the specific 
heat if $N \leq (p+1)$.

The conjectured values (\ref{6.16}), that extend (\ref{6.14}), are shown 
in brackets in Table \ref{table1}. We see a quite good agreement with (\ref{6.16}). 

In the cases where $p=1$ and $M$ odd, like the quantum Ising chain (\ref{HIsing}) and 
the free parafermionic models (\ref{HBaxter}) it was numerically observed that the 
specific heat has a peak in a pseudo-critical point $\tilde{\lambda}_M$ 
, that approaches the true critical point $\lambda_c=1$ as 
$|\tilde{\lambda}_M - \lambda_c| \sim M^{-\nu_{\parallel}}$, with the value 
$\nu_{\parallel} =1$. 
However when we consider the case $p=1$ with $M$ even,
whose quasi-particle energies are given by (\ref{6.12}) for $N=2$, we 
verify that the 
pseudo-critical point approaches the critical point as 
$|\tilde{\lambda}_M - \lambda_c| \sim M^{-1.94}$. For other models with $p>1$, 
the exponents changes as we consider different $\ell$-sequences of lattice 
sizes $M= j\overline{M} + \ell$ ($ j=0,1,\ldots$,$\ell=0,1,\ldots$). 
In fact the finite-size behavior $M^{-\nu_{\parallel}}$ is not a consequence of the FSS theory, and it is not generally expected for open chains 
\cite{BH}. We conjecture that for all the models with any 
$p$ and $N$ we have the simple scaling where $\nu_{\parallel}=1$, as verified in 
the models with $p=1$ and $M$ odd.

\begin{table*}[t] 
\centering
\begin{tabular}{|l|l|l|l|l|l}
\hline \hline
$N$ & $p=2$ & $p=3$ & $p=4$ & $p=5$ \\\hline 
2 & 0.000 [0] & 0.000 [0] & 0.000[0] & 0.000 [0] \\
3 & 0.051 [0] & 0.001 [0]& 0.000 [0]& 0.000 [0]\\
4 & 0.252 [0.25] & 0.092 [0] & 0.000 [0] & 0.000 [0]\\
5 & 0.400 [0.4]& 0.206 [0.2]& 0.002 [0]& 0.000 [0] \\
6 & 0.500 [0.5] & 0.334 [0.333...] & 0.002 [0]& 0.000 [0] \\
7 & 0.571 [0.571...]& 0.428 [0.428...]& 0.287 [0.285...]& 0.153 [0.142...]\\
8 & 0.625 [0.625]& 0.500[0.5] & 0.375 [0.375] & 0.252 [0.25] \\
9 & 0.666 [0.666...]& 0.567 [0.555...]& 0.444 [0.444...]& 0.334 [0.333...]\\
\hline 
$M_{\scriptsize{\mbox{max}}}$ & 2100  &  2560 & 9450 & 10680 \\ \hline \hline
\end{tabular}
\caption{The specific-heat critical exponents $\alpha$ for the 
free-fermionic ($N=2$) 
and free-parafermionic models ($N=3-9$),and parameters $p=2-5$.
The exact predicted values (\ref{6.16}) are shown in brackets. 
It is shown the extrapolated results of the estimator (\ref{6.15}) using 
sequences of lattice sizes up to $M_{\scriptsize{\mbox{max}}}$, shown 
in the last line. }
\label{table1}
\end{table*}

\section{Additional commuting charges and complete set of commuting obervables} 
\label{sec:adic}

Distinct representations of the Hamiltonians (\ref{Hgen}), with $M$ generators 
$\{h_i\}$ satisfying (\ref{halgebra1})-(\ref{halgebra3}), with given values of $p$ and $N$, 
have distinct dimensions. In Section \ref{sec:integrability} we showed the existence of $\overline{M}$ 
commuting charges ($H_M^{(\ell)}, \ell=1,\ldots,\overline{M}$), independently of the 
representation. The eigenvalues of these charges are given in terms of the 
$\overline{M}$ roots of the polynomials $P_M^{(p)}(z)$ and by the set of 
``$Z(N)$ signals" $\{s_1,\ldots,s_{\overline{M}}\} $ (\ref{EHgen}). These are all the possible 
values of the eigenvalues. In the generic case the dimensions of the 
representation  of (\ref{Hgen}) is bigger than $N^{\overline{M}}$, implying degeneracy
in energy as well in all the $\overline{M}$ commuting  charges 
$\{H_M^{(\ell)}\}$. 

Let us consider some free fermionic representations ($N=2$) with $p=1$. For 
$M$ odd the representations of the Ising quantum chain (\ref{HIsing}) has dimension
$2^{(M+1)/2}$ and since in this case $\overline{M}=(M+1)/2$ all the eigenvalues can 
be indexed by the roots of $P_M^{(p)}(z)$ and signals $\{s_i\}$. That is, all 
the $2^{(M+1)/2}$ eigenfunctions are distinctly characterized by the complete 
set of commuting observables (CSCO) $\{H_M^{(\ell)}; \ell=1,\ldots, (M+1)/2\}$.

In the case where $M$ is even, as in (\ref{HIsingeven}), the dimension of the Hilbert 
space is $2^{M/2}$ and since $\overline{M} = M/2$ the charges (\ref{charges}) form 
again a CSCO. This construction is an interesting way to see the fully 
exact integrability of the Ising quantum chains in a finite lattice. 

In the case 
where the free fermionic models are in the representation (\ref{HA1}) with 
dimension $2^M$ the situation is distinct since the number of conserved charges 
$\{H_M^{(\ell)}\}$ is $(M+1)/2$ or $M/2$ if $M$ is odd or even, respectively.
However we can identify an extra set of $Z(2)$ gauge operators $\{g_M^{(i)}\}$:
\bea \label{addcharge1} 
&&\sigma_1^x \sigma_2^z,\; \sigma_3^x\sigma_4^z,\cdots,\;
\sigma_{M-1}^x\sigma_M^z ,\; \sigma_M^x 
\quad (M \,\,\mbox{even}); \nonumber \\
&&\sigma_1^x \sigma_2^z,\; \sigma_3^x\sigma_4^z,\cdots,\;
\sigma_{M-1}^x\sigma_M^z  \quad (M\,\, \mbox{odd}),
\eea
that besides commuting among themselves and with $\{H_M^{(\ell)}\}$ are 
independent. Since this extra set (\ref{addcharge1}) has $(M-1)/2$ or $M/2$ charges, for 
$M$ odd or even, respectively, we have a total of $M$ commuting charges 
forming again a CSCO. Differently from the charges $\{H_M^(\ell)\}$ whose 
eigenvalues are obtained from the roots of the polynomials $P_M^{(p)}(z)$, the 
gauge charges $\{g_M^{(i)}\}$ have eigenvalues $\pm 1$, since $(g_M^{(i)})^2=1$.
The commutation $\left[g_M^{(i)},H_M^{(\ell)}\right]=0$ imply that all the eigenvalues 
of the Hamiltonian (\ref{HA1}), as well as all the charges $\{H_M^{(\ell)}\}$ 
will have, apart from accidental degeneracies, a degeneracy $2^{(M-1)/2}$ or 
$2^M$ for $M$ odd or even, respectively. 

The preceding discussion for $p=1$ and $N=1$ is easily generalized for the 
free parafermionic cases where $N>2$. However, this is not the case for $p>1$. In fact, let us 
consider the fermionic models ($N=2$) with $p=2$ and let us restrict ourselves 
to the representation (\ref{HA}):
\be \label{6.2}
\mathcal{H}_A &=& -\lambda_1\sigma_1^x -\lambda_2\sigma_1^z\sigma_2^x 
-\lambda_3\sigma_1^z\sigma_2^z\sigma_3^x -\lambda_4\sigma_2^z\sigma_3^z\sigma_4^x
 -\non\\ \cdots &-& \lambda_M \sigma_{M-2}^z\sigma_{M-1}^z\sigma_M^x.
\ee
The number of commuting charges $\{H_M^{(\ell)}\}$, whose eigenvalues are given 
in terms of the roots of the polynomial $P_M^{(2)}(z)$ is $2^{\overline{M}}$, while 
the dimension of the representation is $2^M$. We verified by direct 
diagonalizations that for small values of $M$ all the eigenvalues of (\ref{6.2})
has the same degeneracy $2^{M}/2^{\overline{M}}$. As in the case $p=1$ we can also 
identify an extra set $\{g_M^{(i)}\}$ of $Z(2)$ gauge operators:
\beq \label{6.3}
\sigma_1^x\sigma_2^z\sigma_3^z, \; \sigma_4^x\sigma_5^z\sigma_6^z, \; \cdots 
\eeq
forming the commuting set $\{g_M^{(i)},H_M^{(\ell)}\}$. Since there are 
$=\lfloor M/3 \rfloor$ gauge operators and $(g_M^{(i)})^2 =1$, we can explain 
$2^{\tilde{M}}$ of this degeneracy. Although we believe in their existence 
we did not find the extra charges that will complete the CSCO of (\ref{6.2}).

The previous discussions can be easily extended for $N>2$ and $p>2$, and again 
we can explain part of the degeneracies of the Hamiltonian in their 
word representation.

\section{Conclusions} \label{sec:conc}
	
We demonstrated that the general quantum spin chains (\ref{Hgen}) with $M$ 
density operators $\{h_i\}$ satisfying the $p$-exchange algebra 
(\ref{halgebra1})-(\ref{halgebra3}) is exactly integrable. In the bulk limit $M \to \infty$ the 
Hamiltonians belongs to an infinite set of conserved charges. In the generic 
case the number of independent words we can form in the algebra is infinite, 
even for finite $M$. However by including the closure relation (\ref{halgebra3}) in 
the algebra the number of independent words, or the dimension of the Hilbert 
space associated to the Hamiltonian (\ref{Hgen}) is finite for finite $M$. We can 
have then several possible Hamiltonians described for the case $N=2$ in terms 
of spin-$\frac{1}{2}$ Pauli matrices and for $N>2$ by the generalized $Z(N)$ 
Pauli matrizes (\ref{Weyl}). In these cases we show that for arbitrary values of 
the parameter $p$, all the models have a free fermionic ($N=2$) or 
free parafermionic ($N>2$) eigenspectrum. The eigenenergies are given in terms of the 
zeros of the polynomials $P^{(p)}_M(z)$. In the case $p=1$ and 
$\{\lambda_i\}=1$, where the corresponding Hamiltonians are the quantum 
Ising chain ($N=2$) or the $Z(N)$ Baxter parafermionic chain ($N>2$), at their 
critical point, $P^{(p)}_M(z)$ is related to the Chebyshev polynomial of 
second type. We presented several representations of the algebraic Hamiltonian 
(\ref{Hgen}), for several values of $p$. In special we also construct the 
Hamiltonian (\ref{Hgen}) on its word representation, a representation with 
a one-to-one equivalence among the independent words in the algebra and the basis 
vectors spanning the associated Hilbert space.

Although the models, for arbitray $N$ and $p$, have a quite rich phase diagram 
we only consider, in this paper, the isotropic point where all the couplings 
$\{\lambda_i=1\}$ (see (\ref{halgebra3})). We showed that the models are critical at this 
pint for any $p$ and $N$. We calculated at this critical point the exact 
ground-state 
energy for general $p$ and $N$, and it turns out to be  expressed in terms of 
integral representations of Lauricella series.  From our extensive numerical 
studies we conjectured that at this isotropic point the dynamical critical 
exponent is given by $z= (p+1)/N$ and  the specific-heat exponent 
by $\alpha=\max \{0,1-(p+1)/N\}$. This is interesting since most of the known 
critical chains are conformally invariant ($z=1$), and therefore the multispin quantum chains 
provide an excellent lab to understand the universal behavior of the shared 
quantum information measures, like the von Neumann or R\`enyi entanglement 
entropies, in critical quantum chains without an underlying conformal symmetry. The polynomials
$P_M^{(p)}(z)$ may also play an important role in the study of the entanglement 
entropy, see \cite{Crampe2019}.

An interesting direction of investigation is to consider extensions
of the algebra (\ref{halgebra1})-(\ref{halgebra3}) to include
models with periodic boundary conditions. See for example the interesting
recent paper \cite{minami}. In this case, for $p=1$ and $N=2$
the algebra (\ref{halgebra1})-(\ref{halgebra3}) is related to the
Temperley-Lieb and Onsager algebras. For periodic boundary conditions
the eigenspectrum seems to be not a simple 
free particle one, but probably a composition of free particle spectra. As observed
in \cite{Alcaraz_2018} the $Z(N)$ Baxter free parafermionic quantum 
chain, that corresponds to the case $p=1$, has an anomalous behavior for the
ground state energy per site for $N>2$, where the models are non-Hermitian.
It should be also interesting to probe if this anomaly happens for general values of $p$.

We conclude mentioning that although we did not construct the general raising 
and lowering fermionic and parafermionic operators, related to the quantum 
chains (\ref{Hgen}), we believe that this construction should follows the one 
introduced by Fendley \cite{Fendley:2019} for $p=N=2$ exploiting the general 
product formula (\ref{prodaux}).

\appendix

\section{Derivation of the recurrense relation for $\tau_M^{(3)}(u)$.}
\label{appendixA}

In this appendix we derive the recurrence relation (\ref{rectau3}) of Sec. IV.
To simplify the notation let us denote
\be
G_M^{(i)}\equiv G_M^{(i)}(u) = G_M(\omega^i u)\,,\quad i=0,1,2\,,
\ee
so that
\be\label{tau3a}
\tau^{(3)}(u) = G_M^{(0)}G_M^{(1)}G_M^{(2)}\,.
\ee

Using the fundamental relation (\ref{recurrenceformula}) in $G_M^{(i)}$ ($i=0,1,2$)
we obtain,
\be
\tau_M^{(3)}(u)=\tau_{M-1}^{(3)}(u)-u^3h_M^3\tau_{M-(p+1)}^{(3)}(u)
+\Xi_p^{(3)}(u),
\ee
where
\be\label{Xi3}
\Xi_p^{(3)}(u) = A_1(M)u+A_2(M)u^2\,,
\ee
and
\bea \label{A1}
&&A_2(M)=-\mathcal{L}(G_{M-1};G_{M-(p+1)}h_M;G_{M-(p+1)}h_M)_{\omega^2}, 
\nonumber \\
&&A_1(M)=-\mathcal{L}(G_{M-(p+1)}h_M;G_{M-1};G_{M-1})_{\omega}\,,
\eea
where we have introduced the $Z(3)$ cyclic commutator, defined as,
\bea
&&\mathcal{L}(A(u);B(u);C(u))_{\Omega} = A(u)B(\omega u)C(\omega^2u) \nonumber 
\\ &&+
\Omega C(u)A(\omega u)B(\omega^2u)+
\Omega^2 B(u)C(\omega u)A(\omega^2u)\,.
\eea
To proceed it is interesting to define the generalized
operators,
\be\label{A1j}
A_1^{(j)}(M) =
-\mathcal{L}(G_{M-(p+1)}h_M;G_{M-j};G_{M-j})_{\omega},
\ee
and
\bea\label{A2jk}
&&A_2^{(j,k)}(M) =
-\mathcal{L}(G_{M-(p+1)} \nonumber \\
&&;h_{M-j}G_{M-(p+1)-j};h_{M-k}G_{M-(p+1)-k})_{\omega^2}\,.
\ee
We see that $A_1=A_1^{(1)}$.

{\it Recurrence relation for $A_1^{(j)}(M)$}.

 Using (\ref{recurrenceformula})
in both the $G_{M-j}$ of $A_1^{(j)}(M)$ we obtain,
\be \label{A1j}
A_1^{(j)}(M) = \sum_{l=j}^p\left(\gamma_1^{(l)}
+\gamma_2^{(l)}\right)-\sum_{l=j}^p\sum_{k=j}^p\gamma_3^{(l,k)}\,,
\ee
where
\be
\gamma_1^{(l)} = u
\mathcal{L}(G_{M-(p+1)}h_M;h_{M-j}G_{M-(p+1)-j};G_{M-(p+1)})_{\omega} \,,\non
\ee
\be
\gamma_2^{(l)} = u
\mathcal{L}(G_{M-(p+1)}h_M;G_{M-(p+1)};h_{M-j}G_{M-(p+1)-j})_{\omega}\,,\non
\ee
and
\bea
&&\gamma_3^{(l,k)} = u^2
\mathcal{L}(G_{M-(p+1)}h_M \nonumber \\
&&;h_{M-l}G_{M-(p+1)-l};h_{M-k}G_{M-(p+1)-k})_{\omega}\,.
\eea
Expanding $\gamma_1^{(l)}$ and $\gamma_2^{(l)}$ we obtain,
\be \label{g1pg2}
\gamma_1^{(l)}+\gamma_2^{(l)}=-uh_M\omega^2A_1^{(p+1-l)}(M-l)\,,
\nonumber \\
\ee
where
$ l=1,2,\dots,p$,
and  identify,
\be\label{g3simp}
\gamma_3^{(l,k)}=u^2h_MA_2^{(l,k)}(M) \,.
\ee
Using (\ref{it1}) in (\ref{A2jk}) we obtain,
\be\label{A2jj}
A_2^{(j,j)}(M) = -uh_{M-j}^2\omega \sum_{l=0}^{j-1}A_1^{(j-l)}(M-(p+1)-l).
\nonumber
\ee
Using this last expression with (\ref{g1pg2}) and (\ref{g3simp})  
in (\ref{A1j}) we get
\begin{widetext}
\be
A_1^{(j)}(M)&=&-uh_M\omega^2\sum_{l=j}^pA_1^{(p+1-l)}(M-l)
-u^3h_M\omega\sum_{l=j}^ph_{M-l}^2\sum_{l'=0}^{l-1}A_1^{(l-l')}(M-(p+1)-l')
\non\\&-&u^2h_M\sum_{k=j}^p\sum_{l=j+1}^{k-1}\left(A_2^{(l,k)}(M)+A_2^{(k,l)}(M)\right)\,.
\ee
\end{widetext}
Also using (\ref{recurrenceformula}) in (\ref{A2jk}) we obtain
for $j\neq k$,
\be
A_2^{(j,k)}(M)=\alpha_1^{(j,k)}(M)+\alpha_2^{(j,k)}(M)\,,\nonumber
\ee
where
\bea\label{a1jk}
&&\alpha_1^{(j,k)}(M) =
\mathcal{L}(G_{M-(p+1)-j};h_{M-j}G_{M-(p+1)-j} \nonumber \\
&&;h_{M-k}G_{M-(p+1)-k})_{\omega^2}
\,,
\eea
and  for ($j<k$),
\be
\alpha_2^{(j,k)}(M) = -uh_{M-j}\sum_{l=0}^{j-1}A_2^{(k-j,p+1+l-j)}(M-j)\,, \non
\ee
\be
\alpha_2^{(k,j)}(M) = -uh_{M-j}\sum_{l=0}^{j-1}A_2^{(p+1+l-j,k-j)}(M-j)\,. \non
\ee
Combining $\alpha_1^{(j,k)}(M)+\alpha_1^{(k,j)}(M)$ ($j\neq k$) we obtain,
 for $j<k$,
\be
\alpha_1^{(j,k)}(M)+\alpha_1^{(k,j)}(M)=-\omega^2h_{M-j}A_1^{(p+1+j-k)}&&(M-k)\,.
\non
\ee
Then, for $j<k$
\bea\label{A2A2}
A_2^{(j,k)}(M)&+&A_2^{(k,j)}(M)=-\omega^2h_{M-j}A_1^{(p+1+j-k)}(M-k) 
\nonumber \\
&&-uh_{M-j}\sum_{l=0}^{j-1}\left(   A_2^{(p+1+l-j,k-j)}(M-j)\right. 
\nonumber \\
&&\left. +A_2^{(k-j,p+1+l-j)}(M-j) \right).
\eea

{\it Recurrence relation for $A_2(M)$}. 

Using the relation
(\ref{it2}) in $A_2(M)$, given in (\ref{A2A2}), we obtain,
\bea
&&A_2(M) = -uh_M^2\sum_{j=1}^p \nonumber \\
&&\mathcal{L}(G_{M-(p+1)};G_{M-(p+1)}; 
h_{M-j}G_{M-(p+1)-j})_{\omega}\,,
\eea
giving us, from (\ref{A1}),
\be\label{A2n}
A_2(M) = -uh_M^2\omega\sum_{j=1}^pA_1^{(p+1-j)}(M-j)\,.
\ee

Equation (\ref{a1jk}), with the recurrences (\ref{A2A2}) and (\ref{A2n}),
imply that $A_1(M)=A_1^{(1)}$, $A_1^{(l)}(M)$ ($l=1,\dots,M$)
and $A_2(M)$ depend only on the values of $A_1^{(k')}(M-j'),j'>1,k'\leq p,$ and
on $A_2^{(l,l')}(M-j')+A_2^{(l',l)}(M-j),j'>1,1\leq l\neq l'\leq p$, \textit{i.e.}.,
it depends only on the values of $A_1^{(l)}(M')$, $A_2^{(j,k)}(M')$,
evaluated for smaller lattices.

Since $h_M=0$ ($M\leq 0$), $G_M=\id$ ($M\leq 1$) it is simple
 to verify from (\ref{A2A2}) that $A_1^{(0)}(M)=A_2^{(j,k)}(M)=0$
for $M\leq 1$. For $M=2$ we see that $A_1^{(j)}(2)=0$ ($j=1,\dots,p$)
and $A_2^{(j,k)}(2)+A_2^{(k,j)}(2)=0$ ($k\neq j=1,\dots,p$).
From  (\ref{A2A2}) we also see that 
$A_2^{(j,k)}(3)+A_2^{(k,j)}(3)=0$ ($k\neq j=1,\dots,p$),
and from (\ref{a1jk}) we get $A_1^{(l)}$ ($l=1,\dots,p$). Finally
iterating we obtain for (\ref{A1}), (\ref{A2A2}) and (\ref{Xi3}),
\be
A_1(M)=A_2(M)=\Xi_p^{(3)}(u)=0
\ee
for any $M$ and $p$.

This implies that
\be\label{rectaunova}
\tau_M^{(3)}(u)=\tau_{M-1}^{(3)}(u)-u^3h_M^3\tau_{M-(p+1)}^{(3)}(u)\,.
\ee

\section{Eigenspectrum of a quantum chain with $M=5$ sites  and parameter 
$p=3$.} 
\label{appendixB} 
 In this appendix, for the sake of illustration, we give a simple example 
for the quantum chain (\ref{Hgen}) with $M=5$ generators, parameter $p=3$ and $N$ arbitrary.
In this case $\overline{M} = \lfloor (5+3)/4 \rfloor = 2$.

The general Hamiltonian satisfying the algebra (\ref{halgebra1},
\ref{halgebra3}) is given by
\be \label{h1}
\mathcal{H} = -(h_1+h_2+h_3+h_4+h_5).
\ee
There are $2$ conserved charges, i.e., 
\be \nonumber
H^{(1)}_5 = -\mathcal{H}, \quad H^{(2)}_5= h_1 h_5.
\ee
One of the representations of the conserved charges is the 
word representation (\ref{wordrepp}) where
\bea \label{rep1}
\mathcal{H} &=& -\lambda_1X_1 -\lambda_2Z_1X_2 -\lambda_3 Z_1 Z_2 X_3 
-\lambda_4 Z_1Z_2Z_3X_4 \nonumber \\&&- \lambda_5 Z_2Z_3Z_4X_5, \nonumber \\
H_5^{(2)}&=& \lambda_1 \lambda_5 X_1Z_2Z_3Z_4X_5,
\eea
were $X_i,Z_i$ are the $Z(N)$ matrices (\ref{znalgebra}) and the 
coupling constants $\{\lambda_i\}$ are defined by (\ref{halgebra3}).

The fundamental polynomial is obtained by iterating (\ref{recpolN}) or by (\ref{expP})
(compare with Table I, for the case $\lambda_1=\lambda_2=1$):
\be \label{pol1}
P_5^{(3)}(z) = 1 -(\lambda_1^N + \cdots + \lambda_5^N)z +
\lambda_1^N\lambda_5^Nz^2,
\ee
whose roots $z_1$ and $z_2$ give us the quasienergies $\epsilon_1=z_1^{-1/N}$ 
and $\epsilon_2=z_2^{-1/N}$. The predicted $N^2$ eigenvalues of the Hamiltonian and 
 second charge are obtained from (\ref{EHgen}) and (\ref{othercharges}):
\bea \label{levels}
E^{\{s_1,s_2\}}&=& -e^{i\frac{2\pi}{N}s_1}\epsilon_1
- e^{i\frac{2\pi}{N}s_2}\epsilon_2, \nonumber \\
E_2^{\{s_1,s_2\}} &=&e^{i\frac{2\pi}{N}(s_1+s_2)}\epsilon_1 \epsilon_2, 
\nonumber
\eea
where $s_1,s_2=0,1,\ldots,N-1$. In the word representation (\ref{rep1}) the 
hamiltonian has $N^5$ eigenvalues. A direct diagonalization of (\ref{rep1}) 
show us that all levels have the same degenerascy $N^3$.  The ground state 
energy (real for all $N$) is given by $E^{\{0,0\}} = -\epsilon_1 
-\epsilon_2$, while the excited states have complex eigenvalues.

\begin{acknowledgments}
We thank 
 discussions with Jos\'e A. Hoyos and Edu-ardo Novais. The work of FCA was supported in part by the Brazilian agencies FAPESP and CNPq.
RAP was partially supported by FAPESP/CAPES (grant \# 2017/02987-8). 
\end{acknowledgments}

\end{document}